# Kondo effect in an integer-spin quantum dot

S. Sasaki[*], S. De Franceschi[†], J. M. Elzerman[†], W. G. van der Wiel[†], M. Eto[†,‡], S. Tarucha[*,§], and L. P. Kouwenhoven[†]

[*]NTT Basic Research Laboratories, Atsugi-shi, Kanagawa 243-0129, Japan

[†]Department of Applied Physics, DIMES, and ERATO Mesoscopic Correlation Project, Delft University of Technology, PO Box 5046, 2600 GA Delft, The Netherlands

[‡]Faculty of Science and Technology, Keio University, 3-14-1 Hiyoshi, Kohoku-ku, Yokohama 223-8522, Japan

[§] ERATO Mesoscopic Correlation Project, University of Tokyo, Bunkyo-ku, Tokyo 113-0033, Japan

**The Kondo effect is a key many-body phenomenon in condensed matter physics. It concerns the interaction between a localised spin and free electrons. Discovered in metals containing small amounts of magnetic impurities, it is now a fundamental mechanism in a wide class of correlated electron systems[1,2]. Control over single, localised spins has become relevant also in fabricated structures due to the rapid developments in nano-electronics[3,4]. Experiments have already demonstrated artificial realisations of isolated magnetic impurities at metallic surfaces[5,6], nanometer-scale magnets[7], controlled transitions between two-electron singlet and triplet states[8], and a tunable Kondo effect in semiconductor quantum dots[9-12]. Here, we report an unexpected Kondo effect realised in a few-electron quantum dot containing singlet and triplet spin states whose energy difference can be tuned with a magnetic field. This effect occurs for an even number of electrons at the degeneracy between singlet and triplet states. The characteristic energy scale is found to be much larger than for the ordinary spin-1/2 case.**

Quantum dots are small electronic devices[13], which confine a well-defined number of electrons, $N$. The total spin is zero or an integer for $N$ = even and half-integer for $N$ = odd. The latter case constitutes the canonical example for the Kondo effect[14,15] when all electrons can be ignored, except for the one with the highest energy; i.e. the case of a single, isolated spin, $S = 1/2$ (see Fig. 1a). Although the energy level $\varepsilon_o$ is well below the Fermi energies of the two leads, Heisenberg uncertainty allows the electron on the dot to tunnel to one of the leads when it is replaced quickly by another electron. The time scale for such a co-tunneling process[17] is ~$\hbar/U$, where $h = 2\pi\hbar$ is Planck's constant and $U$ is the on-site Coulomb energy. Figure 1a illustrates that particle exchange by co-tunneling can effectively flip the spin on the dot. At low temperature, the coherent superposition of all possible co-tunneling processes involving spin flip can result in a time-averaged spin equal to zero. The whole system, i.e. quantum dot plus electrodes, forms a spin singlet. The energy scale for this singlet state is the Kondo temperature, $T_K$. In terms of density of states, a narrow peak with a width $\sim k_B T_K$ develops at the Fermi energy ($k_B$ is Boltzmann's constant). Note that for $N$ = even and $S$ = 0, co-tunneling gives rise to a lifetime broadening of the confined state, without producing any Kondo resonance. Such even/odd behaviour corresponding to no-Kondo/Kondo has been observed in recent experiments[9,10].

It is also possible that a quantum dot with $N$ = even has a total spin $S$ = 1; e.g. when the last two electrons have parallel spins. If the remaining $N-2$ electrons can be ignored, this corresponds to a triplet state. Parallel spin filling is a consequence of Hund's rule occurring when the gain in exchange energy exceeds the spacing between single-particle states[8]. The spin of the triplet state can also be screened by co-tunneling events. These are illustrated in the center-left side of Fig. 1b. In contrast to single-particle states that are considered in the spin-1/2 Kondo problem, the spin triplet consists of three degenerate two-particle states. Co-tunneling exchanges only one of the two electrons with an electron from the leads. The total spin of the many-body Kondo state depends on how many modes in the leads couple effectively to the dot[18,19]. If there is only one mode, the screening is not complete and the whole system does not reach a singlet state. In this case the Kondo effect is called "underscreened". Calculations show that also for $S$ = 1 a narrow Kondo resonance arises at the Fermi energy, however, the corresponding $T_K$ is typically lower than in the case of $S$ = 1/2 [20,21]. Some experiments have reported the absence of even/odd behaviour[22,23], which may be related to the formation of higher spin states.

Here, we investigate a quantum dot with $N$ = even where the last two electrons occupy a degenerate state of a spin singlet and a spin triplet. Figure 1b illustrates the different co-tunneling processes occurring in this special circumstance. Starting from $|S = 1, S_z = 1\rangle$, where $S_z$ is the $z$-component of the total spin on the dot, co-tunneling via a virtual state $|1/2, 1/2\rangle$, can lead either to the triplet state $|1,0\rangle$, or to the singlet state $|0,0\rangle$. Via a second co-tunneling event the state $|1, -1\rangle$ can be reached. As for the $S$ = 1 case, the local spin can fluctuate by co-tunneling events. By coupling to all triplet states, the singlet state enhances the spin exchange interaction between the dot and the leads, resulting in a higher rate for spin fluctuations. This particular situation yields a strong Kondo effect, which is characterised by an enhanced $T_K$. This type of Kondo effect has not been considered before, probably because a singlet-triplet degeneracy does not occur in magnetic elements. Recent scaling calculations indeed indicate a strong enhancement of $T_K$ at the singlet-triplet degeneracy[24]. Ref. 24 also argues that the total spin of the many-body Kondo state behaves as in the case of $S$ = 1.

Our quantum dot has the external shape of a rectangular pillar (see Fig. 2a,b) and an internal confinement potential close to a two-dimensional ellipse[25]. The tunnel barriers between the quantum dot and the source and drain electrodes are thinner than in our previous devices[8,25] such that co-tunneling processes are enhanced. Figure 2d shows the linear response conductance (dc bias voltage $V_{sd}$ = 0) versus gate voltage, $V_g$, and magnetic field, $B$. Dark blue regions have low conductance and correspond to the regimes of Coulomb blockade for $N$ = 3 to 10. In contrast to previous experiments[9-12] on the Kondo effect, all performed on lateral quantum dots with unknown electron number, here the number of confined electrons is precisely known. Red stripes represent

Coulomb peaks as high as $\sim e^2/h$. The $B$-dependence of the first two lower stripes reflects the ground-state evolution for $N = 3$ and 4. Their similar $B$-evolution indicates that the 3$^{rd}$ and 4$^{th}$ electron occupy the same orbital state with opposite spin, which is observed also for $N = 1$ and 2 (not shown). This is not the case for $N = 5$ and 6. The $N = 5$ state has $S = 1/2$, and the corresponding stripe shows a smooth evolution with $B$. Instead, the stripe for $N = 6$ has a kink at $B \approx 0.22$ T. From earlier analyses[25] and from measurements of the excitation spectrum at finite $V_{sd}$ (discussed below) we can identify this kink with a transition in the ground state from a spin triplet to a spin singlet. Strikingly, at the triplet-singlet transition (at $B = B_0$ in Fig. 2c) we observe a strong enhancement of the conductance. In fact, over a narrow range around 0.22 T, the Coulomb gap for $N = 6$ has disappeared completely.

To explore this conductance anomaly, we show in Fig. 3a differential conductance measurements, d$I$/d$V_{sd}$ vs $V_{sd}$, taken at $B = B_0$ and $V_g$ corresponding to the dotted line in Fig. 2d. At $T = 14$ mK the narrow resonance around zero bias has a full-width-at-half-maximum, FWHM $\approx 30$ μV. This is several times smaller than the lifetime broadening, $\Gamma = \Gamma_R + \Gamma_L \approx 150$ μV, as estimated from the FWHM of the Coulomb peaks. The height of the zero-bias resonance decreases logarithmically with $T$ (see Fig. 3b). These are typical fingerprints of the Kondo effect. From FWHM $\approx k_B T_K$, we estimate $T_K \approx 350$ mK. We note that we can safely neglect the Zeeman spin splitting since $g\mu_B B_0 \approx 5$ μV $<< k_B T_K$, implying that the spin triplet is in fact three-fold degenerate at $B = B_0$. This condition is essential to the Kondo effect illustrated in Fig. 1b. Alternative schemes have recently been proposed for a Kondo effect where the degeneracy of the triplet state is lifted by a large magnetic field[26,27].

For $N = 6$ we find markedly anomalous $T$-dependence only at the singlet-triplet degeneracy. Figure 3c shows the conductance versus $V_g$ for different $T$. The upper panel is at $B = 0.12$ T. The two Coulomb peaks correspond to the transition from $N = 5$ to 6 and from $N = 6$ to 7. The small, short-period modulations superimposed on the Coulomb peaks are due to a weak charging effect in the upper part of GaAs pillar above the dot[28]. We will ignore this fine structure and focus on the general $T$-dependence. Upon increasing $T$, the valley conductance for $N = 6$ goes up due to thermally activated transport. A similar behaviour is seen in the lower graph for $B = 0.32$ T. In contrast, at the singlet-triplet transition for $B = 0.22$ T we find an opposite $T$-dependence, again indicating the formation of a Kondo resonance. At the lowest $T$, the valley conductance is as high as $0.7e^2/h$, which is close to the height of the Coulomb peaks.

The $T$-dependence for $N = 5$ and 7 is visibly different than in the non-Kondo valley for $N = 6$ (lower panel). Such a difference is a manifestation of the ordinary spin-1/2 Kondo effect expected for $N =$ odd. Indeed the corresponding zero-bias resonances are clearly observed (see insets to Fig. 3a). Their height, however, is much smaller than for the singlet-triplet Kondo effect. There is also some indication for a triplet Kondo effect in the $T$-dependence for $N = 6$ at $B = 0.12$ T, although the associated zero-bias anomaly is not as apparent.

We now investigate the effect of lifting the singlet-triplet degeneracy by changing $B$ at a fixed $V_g$ corresponding to the dotted line in Fig. 2d. Near the edges of this line, i.e. away from $B_0$, the Coulomb gap is well developed as denoted by the dark colours. The d$I$/d$V_{sd}$ vs $V_{sd}$ traces still exhibit anomalies, however, now at finite $V_{sd}$ (see Fig. 4a). For $B = 0.21$ T we observe the singlet-triplet Kondo resonance at $V_{sd} = 0$. At higher $B$ this resonance splits apart showing two peaks at finite $V_{sd}$. It is important to note that these peaks occur inside the Coulomb gap. They result from "inelastic" co-tunneling events[17,29] where "inelastic" refers to exchanging energy between quantum dot and electrodes (see right drawing in Fig. 4d). The upper traces in Fig. 4a, for $B < 0.21$ T, also show peak structures, although less pronounced.

In Fig. 4b we plot the positions of the d$I$/d$V_{sd}$ peaks in the plane of $V_{sd}$ and $B$. The symbols refer to different gate voltages indicating that these positions do not depend on $V_g$. The solid lines are obtained from the excitation spectrum measured (see Fig. 4c) in direct tunneling (explained on the left in Fig. 4d). They represent the measured $B$-dependence of the singlet-triplet energy difference, $\Delta$. The fact that these independent measurements coincide implies that inelastic co-tunneling occurs when $eV_{sd} = \pm\Delta$. Note that this condition is independent of $V_g$, consistently with our observation. We believe that for small $\Delta$ ($\lesssim k_B T_K$) the split resonance reflects the singlet-triplet Kondo anomaly shifted to finite bias. This resembles the splitting of the Kondo resonance by the Zeeman effect[9,10,30], although on a very different $B$-scale. In the present case, the splitting occurs between two different multi-particle states and originates from the $B$-dependence of the orbital motion. For increasing $\Delta$, the shift to larger $V_{sd}$ induces spin-decoherence processes, which broaden and suppress the finite-bias peaks[30]. For $B \approx 0.39$ T the peaks have evolved into steps[29] which may indicate that the spin-coherence associated with the Kondo effect has completely vanished.

**Acknowledgements.** We thank Yu. V. Nazarov, K. Maijala, S. M. Cronenwett, J. E. Mooij, and Y. Tokura for discussions. We acknowledge financial support from the Specially Promoted Research, Grant-in-Aid for Scientific Research, from the Ministry of Education, Science and Culture in Japan, from the Dutch Organisation for Fundamental Research on Matter (FOM), from the NEDO joint research program (NTDP-98), and from the EU via a TMR network.

**Correspondence and requests for materials should be addressed to S. D. F. (e-mail: silvano@qt.tn.tudelft.nl).**

**Figure 1.** Spin-flip processes leading to ordinary and singlet-triplet Kondo effect in a quantum dot. **a**, Co-tunneling event in a spin-1/2 quantum dot for $N$ = odd. Only the highest-energy electron is shown occupying a single spin-degenerate level, $\varepsilon_o$. (The case of two, or more, closely spaced levels has also been considered theoretically within the context of the spin-1/2 Kondo effect[16].) The green panels refer to $S_z$ = 1/2 and −1/2 ground states, which are coupled by a co-tunneling event. The two tunnel barriers have tunneling rates $\Gamma_R$ and $\Gamma_L$. In the Coulomb blockade regime ($|\varepsilon_o| \sim U$) adding or subtracting an electron from the dot implies an energy cost $\sim U$. Hence the intermediate step (orange panel) is a high-energy, virtual state. The spin-flip event depicted here is representative of a large number of higher-order processes which add up coherently such that the local spin is screened. This Kondo effect leads to an enhanced linear-response conductance at temperatures $T \lesssim T_K$. **b**, Co-tunneling in an integer-spin quantum dot for $N$ = even at a singlet-triplet degeneracy. Two electrons can share the same orbital with opposite spins (singlet state in the blue panel) or occupy two distinct orbitals in one of the three spin-triplet configurations (green panels). The different spin states are coupled by virtual states (orange panels). Similar to the spin-1/2 case, spin-flip events can screen the local magnetic moment. Note that an $S$ = 1 Kondo effect only involves $|1,+1\rangle$, $|1,0\rangle$, and $|1,−1\rangle$.

**Figure 2.** Sample description, energy spectrum and magnetic-field evolution of the ground state. **a**, Cross-section of our rectangular quantum dot. The semiconductor material consists of an undoped AlGaAs(7nm)/InGaAs(12nm)/AlGaAs(7nm) double barrier structure sandwiched between n-doped GaAs source and drain electrodes. A gate electrode surrounds the pillar and is used to control the electrostatic confinement in the quantum dot. A dc bias voltage, $V_{sd}$, is applied between source and drain and current, $I$, flows vertically through the pillar. In addition to $V_{sd}$, we apply a modulation with rms amplitude $V_{ac}$ = 3 μV at 17.7 Hz for lock-in detection. The gate voltage, $V_g$, can change the number of confined electrons, $N$, one-by-one from ~10 at $V_g$ = 0 to 0 at $V_g$ = −1.8 V. A magnetic field, $B$, is applied along the vertical axis. Temperature, $T$, is varied between 14 mK and 1 K. Our lowest effective electron temperature is 25±5 mK. **b**, Scanning electron micrograph of a quantum dot with dimensions 0.45×0.6 μm$^2$ and height of ~0.5 μm. **c**, Schematic energy spectrum. Solid lines represent the $B$-evolution of the first four orbital levels in a single-particle model. The dashed line is obtained by subtracting the two-electron exchange coupling from the fourth level. At the crossing between this dashed line and the third orbital level at $B = B_0$ the ground state for $N$ = 6 undergoes a triplet-to-singlet transition. $B_0 \approx 0.22$ T with a slight dependence on $V_g$. We define $\Delta$ as the energy difference between the triplet and the singlet states. **d**, Color-scale representation of the linear conductance versus $V_g$ and $B$. Red stripes denote conductance peaks of height $\sim e^2/h$. Blue regions of low conductance indicate Coulomb blockade. The $V_g$-position of the stripes reflects the ground state evolution with $B$, for $N$ = 3 to 10. The $N$ = 6 ground state undergoes a triplet-to-singlet transition at $B_0 \approx 0.22$ T, which results in a conductance anomaly inside the corresponding Coulomb gap.

**Figure 3.** Zero-bias resonance and *T*-dependence of the conductance at the singlet-triplet degeneracy. **a,** Kondo resonance at the singlet-triplet transition. The d*I*/d$V_{sd}$ vs $V_{sd}$ curves are taken at $V_g$ = -0.72 V, *B* = 0.21 T and for *T* = 14, 65, 100, 200, 350, 520, and 810 mK. Insets to **a**: Kondo resonances for *N* = 5 (left inset) and *N* = 7 (right inset), measured at $V_g$ = –0.835 V and $V_g$ = –0.625 V, respectively, and for *B* = 0.11 T and T = 14 mK. **b,** Peak height of zero-bias Kondo resonance *vs T* as obtained from **a** (solid diamonds). The line demonstrates a logarithmic *T*-dependence, which is characteristic for the Kondo effect. The saturation at low *T* is likely due to electronic noise. **c,** *T*-dependence of the linear conductance versus $V_g$ for *B* = 0.12 T (spin-triplet ground state), *B* = 0.22 T (singlet-triplet degeneracy), and *B* = 0.32 T (spin-singlet ground state). Each panel shows 7 traces at *T* = 20 (dot-dashed line), 35, 70, 120, 260, 490 (solid lines), 1050 (dashed line) mK. The arrows emphasize the temperature dependence in the valley for *N* = 6.

**Figure 4.** Singlet-triplet energy separation tuned by a magnetic field. **a,** d*I*/d$V_{sd}$ vs $V_{sd}$ characteristics taken along the dotted line in Fig. 2d ($V_g$ = –0.72 V) at equally spaced magnetic fields *B* = 0.11,0.13,...,0.39 T. Curves are offset by 0.25 $e^2$/h. **b,** Extracted peak positions from d*I*/d$V_{sd}$ curves in **a** versus *B*. Each symbol refers to a gate voltage taken between –0.72 and –0.70 V. **c,** Color-scale plot of d*I*/d$V_{sd}$ measured at $V_{sd}$ = 0.67 mV as a function of *B* and $V_g$. The solid line identifies the ground state whereas the dashed line indicates the first excited state[8] for *N* = 6. From their difference we extract the singlet-triplet energy splitting, $\Delta(B)$, using a proportionality factor $\alpha$ =6.7 meV/V to convert gate voltage into energy. The two solid lines in **b** represent ±$\Delta(B)$ with a horizontal shift of 0.08 T to compensate for the shift of the singlet-triplet transition to a higher magnetic field in a high-bias measurement. **d,** Energy diagrams for two different transport regimes, both with $eV_{sd}$ = $\Delta$. Left: both ground and excited state lie between the two Fermi energies, so two channels are available for direct tunneling. The excitation spectrum in (c) is measured in this regime. Right: both ground and excited state lie below the Fermi levels of the leads (Coulomb blockade regime). Inelastic co-tunneling is illustrated where one electron tunnels out of the lower energy state and another tunnels into the higher energy state.

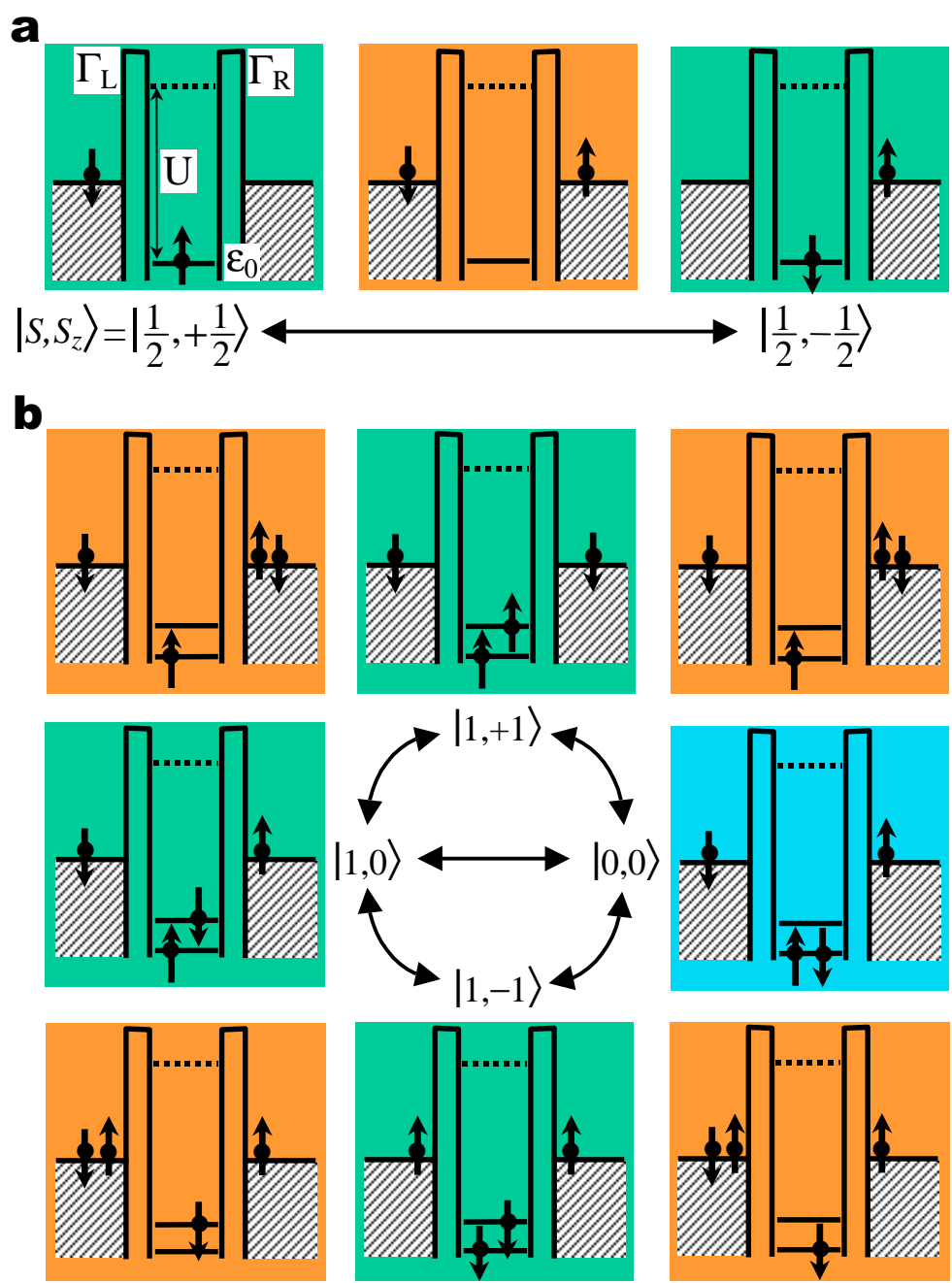

Fig. 1

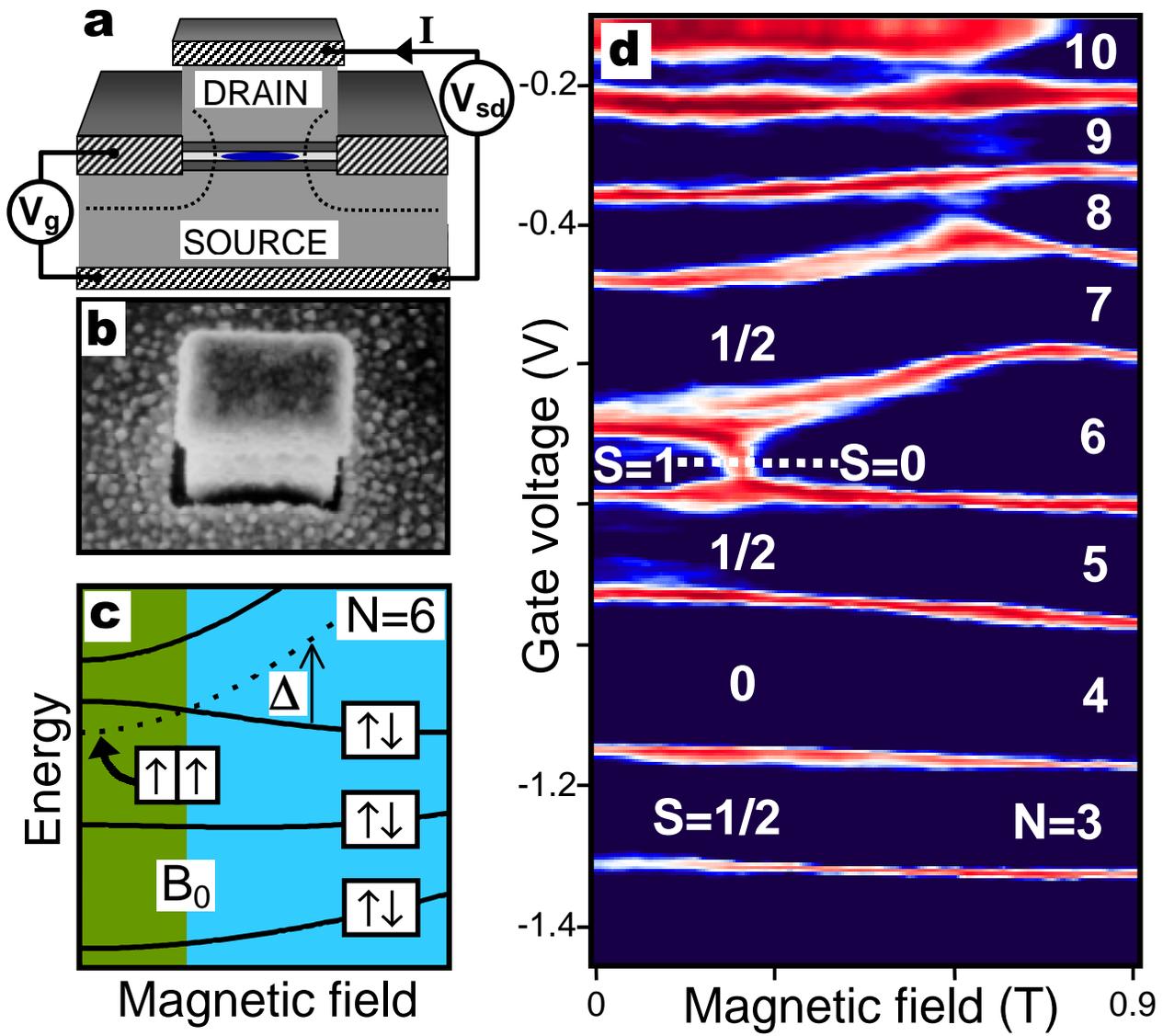

Fig. 2

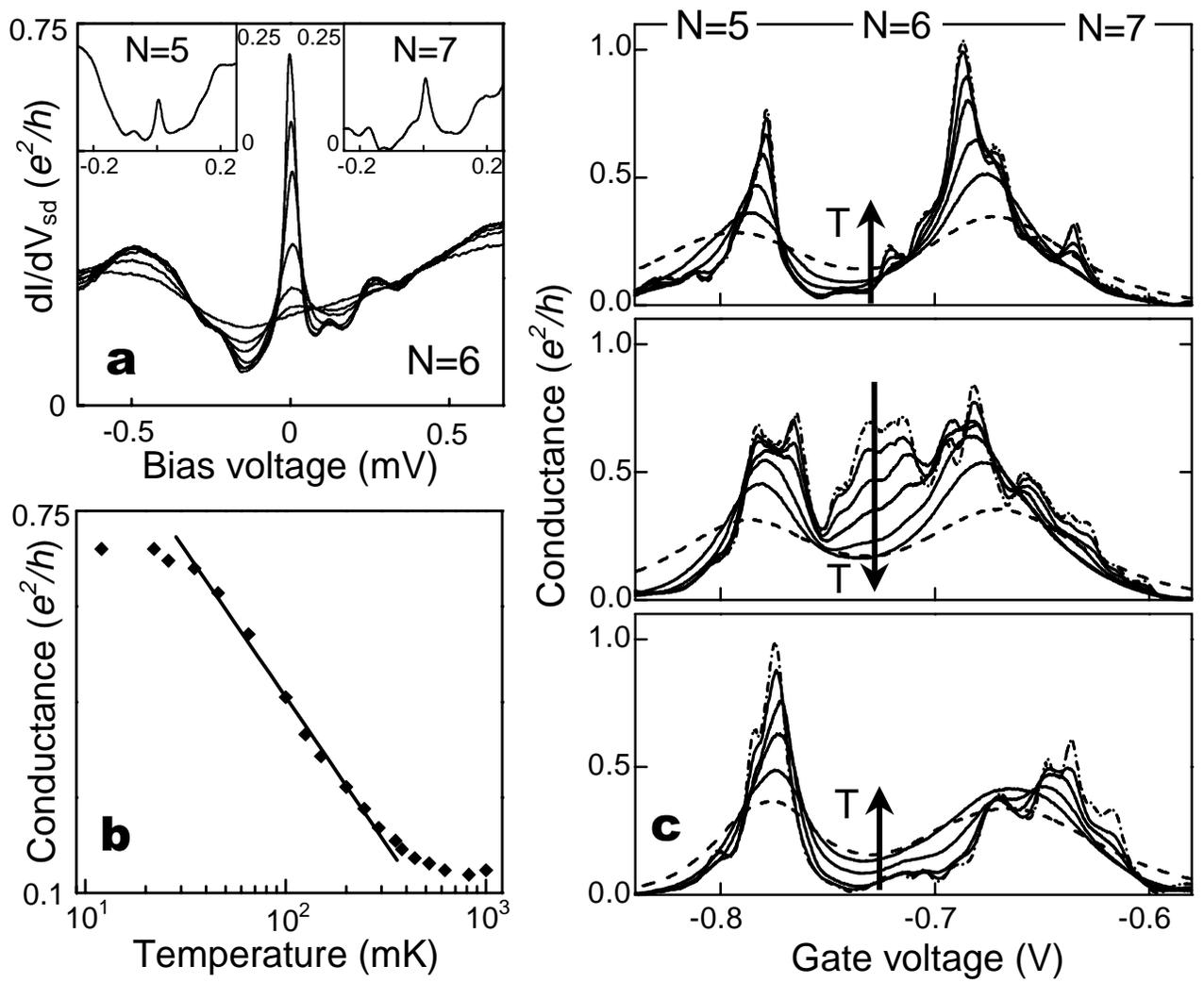

Fig. 3

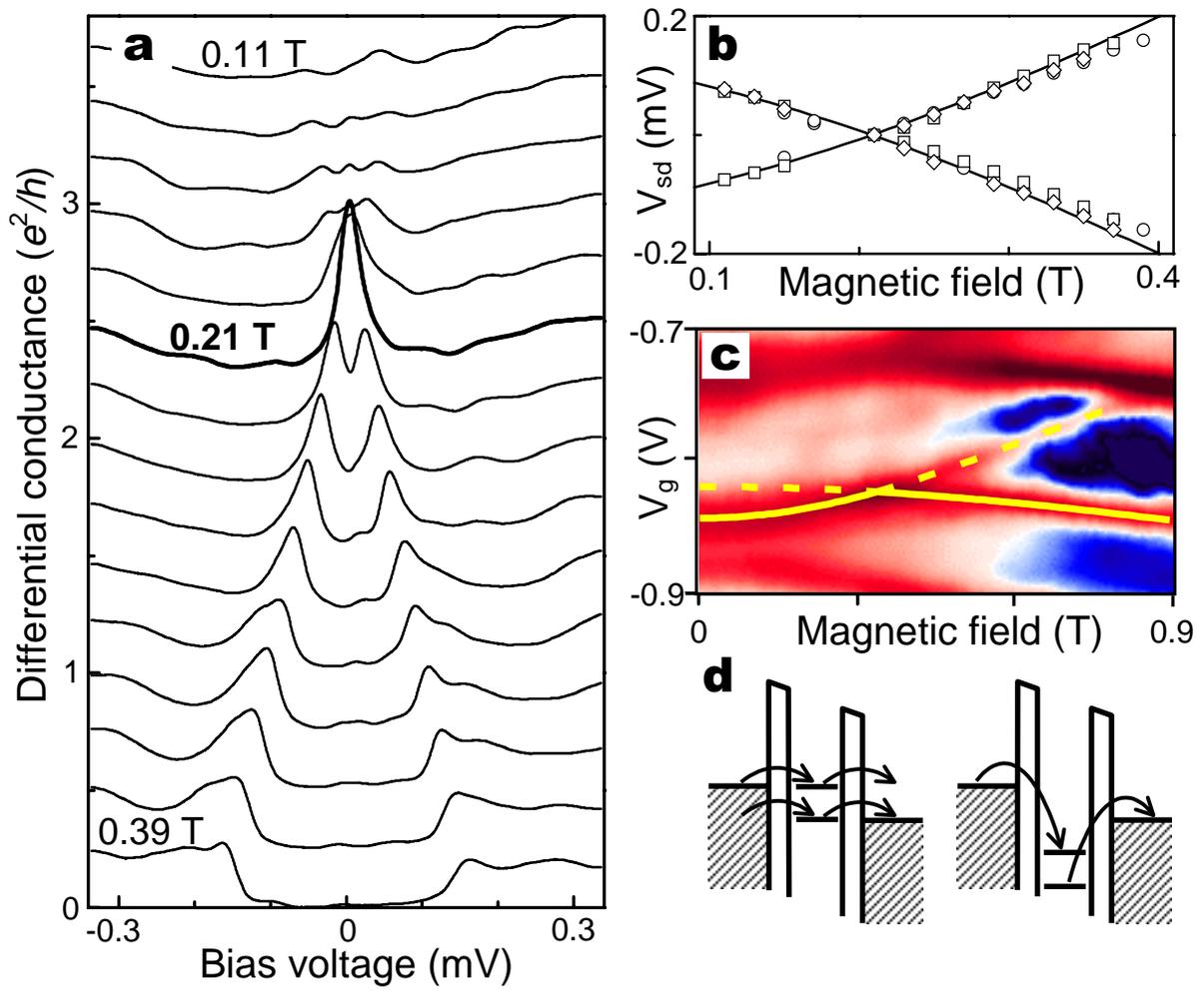

Fig. 4